\documentclass{elsart}
\usepackage{epsfig}
\usepackage{amsmath}
\usepackage{color}
\usepackage{amssymb}
\usepackage{graphicx}
\usepackage{subfigure}

\newcommand \be{\begin{equation}}
\newcommand \ba{\begin{eqnarray}}
\newcommand \ee{\end{equation}}
\newcommand \ea{\end{eqnarray}}
\begin{document}
\runauthor{Zhou and Sornette} \markboth{A}{B}
\begin{frontmatter}
\title{Lead-lag cross-sectional structure and
detection of correlated-anticorrelated regime shifts: application to the
volatilities of inflation and economic growth rates}
\author[ecust,nice]{\small{Wei-Xing Zhou}},
\author[nice,ETH]{\small{Didier Sornette}\thanksref{EM}}
\address[ecust]{School of Business and Research Center of Systems
Engineering, East China University of Science and Technology,
Shanghai 200237, China}
\address[ETH]{Department of Management, Technology
and Economics, ETH Zurich\\ CH-8032 Zurich, Switzerland}
\address[nice]{Laboratoire de Physique de la Mati\`ere Condens\'ee,
CNRS UMR 6622 and Universit\'e de Nice-Sophia Antipolis, 06108 Nice
Cedex 2, France}
\thanks[EM]{Corresponding author. {\it E-mail address:}\/
sornette@ethz.ch (D. Sornette)\\
http://www.er.ethz.ch/}

\begin{abstract}
We have recently introduced the ``thermal optimal path'' (TOP)
method to investigate the real-time lead-lag structure between two
time series. The TOP method consists in searching for a robust
noise-averaged optimal path of the distance matrix along which the
two time series have the greatest similarity. Here, we generalize
the TOP method by introducing a more general definition of distance
which takes into account possible regime shifts between positive and
negative correlations. This generalization to track possible changes
of correlation signs is able to identify possible transitions from
one convention (or consensus) to another. Numerical simulations on
synthetic time series verify that the new TOP method performs as
expected even in the presence of substantial noise. We then apply it
to investigate changes of convention in the dependence structure
between the historical volatilities of the USA inflation rate and
economic growth rate. Several measures show that the new TOP method
significantly outperforms standard cross-correlation methods.
\end{abstract}


\end{frontmatter}

\typeout{SET RUN AUTHOR to \@runauthor}
%

\section{Introduction}
\label{s1:intro}

The study of the lead-lag structure between two time series $X(t)$
and $Y(t)$ has a long history, especially in economics, econometrics
and finance, as it is often asked which economic variable might
influence other economic phenomena. A simple measure is the lagged
cross-correlation function $C_{X,Y}(\tau)=\langle X(t) Y(t+\tau)
\rangle / \sqrt{{\rm Var}[X] {\rm Var}[Y]}$, where the brackets
$\langle x \rangle$ denotes the statistical expectation of the
random variable $x$ and ${\rm Var}[x]$ is the variance of $x$. The
observation of a maximum of $C_{X,Y}(\tau)$ at some non-zero
positive time lag $\tau$ implies that the knowledge of $X$ at time
$t$ gives some information on the future realization of $Y$ at the
later time $t+\tau$. However, such correlations do not imply
necessarily causality in a strict sense as a correlation may be
mediated by a common source influencing the two time series at
different times. The concept of Granger causality bypasses this
problem by taking a pragmatic approach based on predictability: if
the knowledge of $X(t)$ and of its past values improves the
prediction of $Y(t+\tau)$ for some $\tau>0$, then it is said that
$X$ Granger causes $Y$ (see, e.g.,
\cite{Granger-1980-JEDC,Ashley-Granger-Schmalensee-1980-Em,Engle-White-1999}).
Such a definition does not address the fundamental philosophical and
epistemological question of the real causality links between $X$ and
$Y$ but has been found useful in practice. Our approach is similar
in that it does not address the question of the existence of a
genuine causality but attempts to detect a dependence structure
between two time series at non-zero (possibly varying) lags. We thus
use the term ``causality'' in a loose sense embodying the notion of
a dependence between two time series with a non-zero lag time.

Many alternative methods have been developed in the physical
community. Quiroga et al. proposed a simple and fast method to
measure synchronicity and time delay patterns between two time
series based on event synchronization
\cite{Quiroga-Kreuz-Grassberger-2002-PRE}. Furthermore, as a
generalization of the concept of recurrence plot to analyze complex
chaotic time series \cite{Eckmann-Kamphorst-Ruelle-1987-EPL}, Marwan
et al. developed cross-recurrence plot based on a distance matrix to
unravel nonlinear mapping of times between two systems
\cite{Marwan-Kurths-2002-PLA,Marwan-Thiel-Nowaczyk-2002-NPG}. In
Ref.~\cite{Sornette-Zhou-2005-QF}, we have introduced a novel
non-parametric method to test for the dynamical time evolution of
the lag-lead structure between two arbitrary time series based on a
thermal averaging of optimal paths embedded in the distance matrix
previously introduced in cross-recurrence plots. This method ignores
the thresholds used previously in constructing cross recurrence plot
\cite{Marwan-Kurths-2002-PLA,Marwan-Thiel-Nowaczyk-2002-NPG} and
focuses on the distance matrix. The idea consists in constructing a
distance matrix based on the matching of all sample data pairs
obtained from the two time series under study. The lag-lead
structure is searched for as the optimal path in the distance matrix
landscape that minimizes the total mismatch between the two time
series, and that obeys a one-to-one causal matching condition. To
make the solution robust with respect to the presence of noise that
may lead to spurious structures in the distance matrix landscape,
Sornette and Zhou generalized this search for a single absolute
optimal path by introducing a fuzzy search consisting in sampling
over all possible paths, each path being weighted according to a
multinomial logit or equivalently Boltzmann factor proportional to
the exponential of the global mismatch of this path
\cite{Sornette-Zhou-2005-QF}. The method is referred to in the
sequel as the thermal optimal path (TOP). Zhou and Sornette
investigated further the TOP method by considering difference
topologies of feasible paths and found that the two-layer scheme
gives the best performance \cite{Zhou-Sornette-2006-JMe}.

Here, we generalize the TOP method by introducing a definition of
distance which takes into account possible regime shifts between
positive and negative correlations. This extension allows us to
detect possible changes in the sign of the correlation between the
two time series. This is in part motivated by the problem of
identifying changes of conventions in economic and financial time
series. Keynes \cite{Keynes-1936} and Orl\'ean
\cite{Orlean-1986-Ec,Orlean-1987-Ca,Orlean-1989-RE,Orlean-1992-JEE,Orlean-2004,Boyer-Orlean-2004,Orlean-2004-Ra}
developed the concept of convention, according to which a pattern
can emerge from the self-fulfilling belief of agents acting on the
belief itself. Conventions are subject to shifts: in a recent study,
Wyart and Bouchaud claimed that the correlation between bond markets
and stock markets was positive in the past (because low long term
interest rates should favor stocks), but has recently quite suddenly
become negative as a new ``Flight To Quality'' convention has set
in: selling risky stocks and buying safe bonds has recently been the
dominant pattern \cite{Wyart-Bouchaud-2006-JEBO}. Similarly, Liu and
Liu analyzed the nexus between the historical volatility of the
output and of the inflation rate, using Chinese data from 1992 to
2004 \cite{Liu-Liu-2005-ERJ}. They found that there is a strong
correlation between the two volatilities and, what is more
interesting, that the rolling correlation coefficient changes sign.
Such a change of sign of the correlation may be attributed either to
a shift in convention and/or to changing macroeconomic variables,
the two being possible entangled. Our method does not address the
source of the change of the sign of the correlation but provides
nevertheless a preliminary tool for detecting such changes of
correlations in an time-adaptive lead-lag framework.

The paper is organized as follows. In Section \ref{s1:Top}, we
present a brief description of our generalized TOP method.  We
recall that an advantage of the  TOP method is that it does not
require any {\it{a priori}} knowledge of the underlying dynamics.
The new TOP method is illustrated with the help of synthetic
numerical simulations in Section \ref{s1:NumSim}. Section
\ref{s1:Appl} presents the application of the method to the
investigation of a possible change of dependence between the
historical volatility of the USA inflation rate and the economic
growth rate. Section \ref{s1:concl} concludes.

\section{Thermal optimal path method \label{s1:Top}}

In Refs.\cite{Sornette-Zhou-2005-QF,Zhou-Sornette-2006-JMe}, we have
presented the TOP method and several tests and applications. In this
section, to be self-contained, we briefly recall its main
characteristics in the context of the new proposed distance.

Consider two standardized time series $\{X(t_1):t_1=0,...,N\}$ and
$\{Y(t_2):t_2=0,...N\}$. The elements of the distance matrix
$E_{X,Y}$ between $X$ to $Y$ used in
Refs.~\cite{Sornette-Zhou-2005-QF,Zhou-Sornette-2006-JMe} are
defined as
\begin{equation}
\epsilon_-(t_1,t_2) = [X(t_1)-Y(t_2)]^2~. \label{Eq:DM:minus}
\end{equation}
The value $[X(t_1)-Y(t_2)]^2$ defines the distance between the
realizations of the first time series at time $t_1$ and the second
time series at time $t_2$.

The distance matrix (\ref{Eq:DM:minus}) tracks the co-monotonic
relationship between $X$ and $Y$. But, two time series can be more
anti-monotonic than monotonic, i.e., they tend to take opposite
signs. Consider two limiting cases: (i) $Y(t)=X(t)$ and (ii)
$Y(t)=-X(t)$. Obviously, using the traditional distance
(\ref{Eq:DM:minus}) identifies case (i) as minimizing expression
(\ref{Eq:DM:minus}) for $t_1=t_2$ (actually the minimum is
identically zero in this special case). In contrast, notwithstanding
the fact that $Y(t)$ is perfectly (anti-)correlated with $X(t)$, the
naive idea of minimizing the distance (\ref{Eq:DM:minus}) between
the two time series becomes meaningless. In order to diagnose the
occurrence of anti-correlation, one needs to consider the
``anti-monotonic'' distance
\begin{equation}
\epsilon_{+}(t_1,t_2) = [X(t_1)+Y(t_2)]^2~. \label{Eq:DM:plus}
\end{equation}
The $+$ sign ensures a correct search of synchronization between two
anti-correlated time series. More generally, $X$ and $Y$ may exhibit
more complicated lead-lag correlation relationships, positive
correlation over some time intervals and negative correlation at
other times (as in the change of conventions mentioned in the
introduction). In order to address all possible situations, we
propose to use the mixed distance expressed as follows:
\begin{equation}
\epsilon_{\pm}(t_1,t_2) =
\min[\epsilon_{-}(t_1,t_2),\epsilon_{+}(t_1,t_2)]~. \label{Eq:DM:pm}
\end{equation}

\begin{figure}[htb]
\centering
\includegraphics[width=9cm]{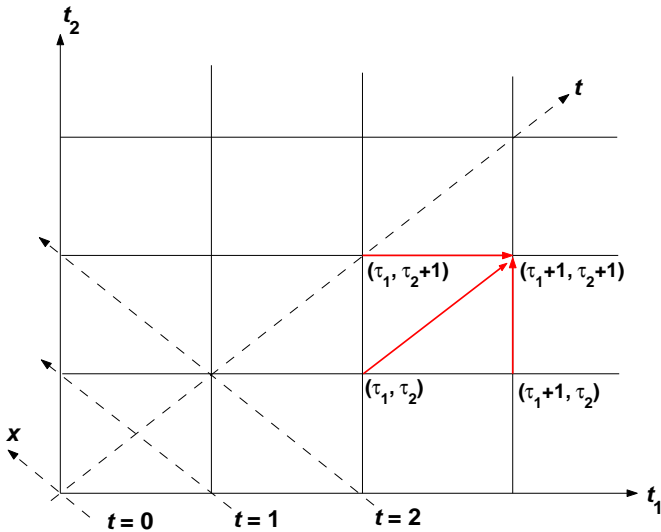}
\caption{(Color online) Representation of the two-layer approach in
the lattice $(t_1,t_2)$ and of the rotated frame $(t,x)$ as defined
in the text. The three arrows depict the three moves that are
allowed to reach any node in one step. } \label{Fig:TOP:TMM}
\end{figure}

Fig.~\ref{Fig:TOP:TMM} is a schematic representation of how lead-lag
paths are defined. The first (resp. second) time series is indexed
by the time $t_1$ (resp. $t_2$). The nodes of the plane carry the
values of the distance (\ref{Eq:DM:pm}) for each pair $(t_1,t_2)$.
The path along the diagonal corresponds to taking $t_1=t_2$, i.e.,
compares the two time series at the same time. Paths below (resp.
above) the diagonal correspond to the second time series lagging
behind (resp. leading) the first time series. The figure shows three
arrows which define the three causal steps (time flows from the past
to the future both for $t_1$ and $t_2$) allowed in our construction
of the lead-lag paths. A given path selects a contiguous set of
nodes from the lower left to the upper right. The relevance or
quality of a given path with respect to the detection of the
lead-lag relationship between the two time series is quantified by
the sum of the distances (\ref{Eq:DM:pm}) along its length.

As shown in the figure, it is convenient to use the rotated
coordinate system $(x,t)$ such that
\begin{equation}
\left\{
   \begin{array}{ccl}
    t_1 &=& 1+\left(t-x\right)/2~ \\
    t_2 &=& 1+\left(t+x\right)/2~
    \end{array}
\right., \label{Eq:AxesTransform2}
\end{equation}
where $t$ is in the main diagonal direction of the $(t_1,t_2)$
system and $x$ is perpendicular to $t$. The origin $(x=0,t=0)$
corresponds to $(t_1=1,t_2=1)$. Then, the standard reference path is
the diagonal of equation $x=0$, and paths which have $x(t) \neq 0$
define varying lead-lag patterns. The idea of the TOP method is to
identify the lead-lag relationship between two time series as the
best path in a certain sense. One could first infer that the best
path is the one which has the minimum sum of its distances
(\ref{Eq:DM:pm}) along its length (paths are constructed with equal
lengths so as to be comparable). The problem with this idea is that
the noises decorating the two time series introduce spurious
patterns which may control the determination the path which
minimizes the sum of distances, leading to incorrect inferred
lead-lag relationships. In
Refs.\cite{Sornette-Zhou-2005-QF,Zhou-Sornette-2006-JMe}, we have
shown that a robust lead-lag path is obtained by defining an average
over many paths, each weighted according to a Boltzmann-Gibbs
factor, hence the name ``thermal'' optimal path method.

Concretely, we first calculate the partition functions $G(x,t)$ and
their sum $G(t)=\sum_x G(x,t)$ so that $G(x,t)/G(t)$ can be interpreted as the
probability for a path to be at distance $x$ from the diagonal for a
distance $t$ along the diagonal. This probability $G(x,t)/G(t)$ is determined as
a compromise between minimizing the mismatch (similar to an ``energy'') and maximizing
the combinatorial weight of the number of paths with similar mismatchs in
a neighborhood (similar to an ``entropy''). As illustrated in Figure
\ref{Fig:TOP:TMM}, in order to arrive at $(t_1+1, t_2+1)$, a path
can come from $(t_1+1, t_2)$ vertically, $(t_1, t_2+1)$
horizontally, or $(t_1, t_2)$ diagonally. The recursive equation on
$G(x,t)$ is therefore
\begin{equation}\label{Eq:RecurG:xt}
      G(x,t+1) = [G(x-1,t)+ G(x+1,t)+G(x,t-1)]e^{-\epsilon_{\pm}(x,t)/T}~,
\end{equation}
where $\epsilon_{\pm}(x,t)$ is defined by (\ref{Eq:DM:pm}). This
recursion relation uses the same principle and is derived following
following the work of Wang et al.
\cite{Wang-Havlin-Schwartz-2000-JPCB}. To $G(x,t)$ at the $t$-th
layer, we need to know and bookkeep the previous two layers from
$G(\cdot,t-2)$ to $G(\cdot,t-1)$. After $G(\cdot,t)$ is determined,
the $G$'s at the two layers are normalized by $G(t)$ so that
$G(x,t)$ does not diverge at large $t$. We stress that the boundary
condition of $G(x,t)$ plays an crucial role. For $t=0$ and $t=1$,
$G(x,t) = 1$. For $t>1$, the boundary condition is taken to be
$G(x=\pm t,t) = 0$, in order to prevent paths to remain on the
boundaries.

Once the partition functions $G(x,t)$ have been calculated, we can
obtain any statistical average related to the positions of the paths
weighted by the set of $G(x,t)$. For instance, the local time lag
$\langle{x(t)}\rangle$ at time $t$ is given by
\begin{equation}
    \langle{x(t)}\rangle = \sum_x {xG(x,t)/G(t)}~.
    \label{Eq:Xave}
\end{equation}
Expression (\ref{Eq:Xave}) defines $\langle{x}\rangle$(t) as the
thermal average of the local time lag at $t$ over all possible
lead-lag configurations suitably weighted according to the
exponential of minus the measure $\epsilon_{\pm}(x,t)$ of the
similarities of two time series. For a given $x_0$ and temperature
$T$, we determine the thermal optimal path $\langle{x}\rangle(t)$.
We can also define an ``energy'' $e_T(x_0)$ to this path, defined as
the thermal average of the measure $\epsilon_{\pm}(x,t)$ of the
similarities of two time series:
\begin{equation}\label{Eq:e}
        e_T(x_0) = \frac{1}{2(N-|x_0|)-1}\sum_{t=|x_0|}^{2N-1-|x_0|}
        \sum_x {\epsilon_{\pm}(x,t)G(x,t)/G(t)}~.
\end{equation}
Obviously, the same set of calculations can be performed with
$\epsilon_-$ given by (\ref{Eq:DM:minus}) or with $\epsilon_{+}$
given by (\ref{Eq:DM:plus}). The former case has been investigated
in Refs.\cite{Sornette-Zhou-2005-QF,Zhou-Sornette-2006-JMe}.

\section{Numerical experiments of the TOP approach on synthetic examples}
\label{s1:NumSim}

We now present synthetic tests of the efficiency of the optimal
thermal causal path method to detect multiple changes of regime.
Consider the following model
\begin{equation}
Y(t)=\left\{
\begin{array}{lr}
     +X(t-10) + \eta,  & ~~~~1\le t \le 100\\
     -X(t-~5) + \eta,  & ~~101\le t \le 200\\
     +X(t+~5) + \eta,  & ~~201\le t \le 300\\
\end{array}
\right.~, \label{Eq:Jump}
\end{equation}
where $\eta$ is a Gaussian white noise with variance $\sigma_\eta^2$
and zero mean. By construction, the time series $Y$ is lagging
behind $X$ with $\tau = 10$ in the first $100$ time steps, $Y$ is
still lagging behind $X$ with a reduced lag $\tau = 5$ in the next
$100$ time steps, and finally $Y$ leads $X$ with a lead time
$\tau=-5$ in the last $100$ time steps. In addition, $Y$ becomes
negatively correlated with $X$ in the middle interval, while it is positively
correlated with $X$ in the first and third interval.
The time series $X$ is
assumed to be the first-order auto-regressive process
\begin{equation}\label{Eq:TOP:AR}
    X(t) = 0.7X(t-1) + \xi~
\end{equation}
where $\xi$ is an i.i.d. white noise with zero mean and variance
$\sigma_\xi^2$. Our results are essentially the same when $X$ is
itself a white noise process. The two time series are standardized
before the construction of the distance matrix. Therefore, there is
only one parameter $f\triangleq\sigma_\xi/\sigma_\eta$
characterizing the signal-over-noise ratio of the lead-lag
relationship between $X$ and $Y$. We use $f=1/5$ in the simulations
presented below, corresponding to a weak signal-to-noise ratio.

\begin{figure}[htb]
\centering
\includegraphics[width=9cm]{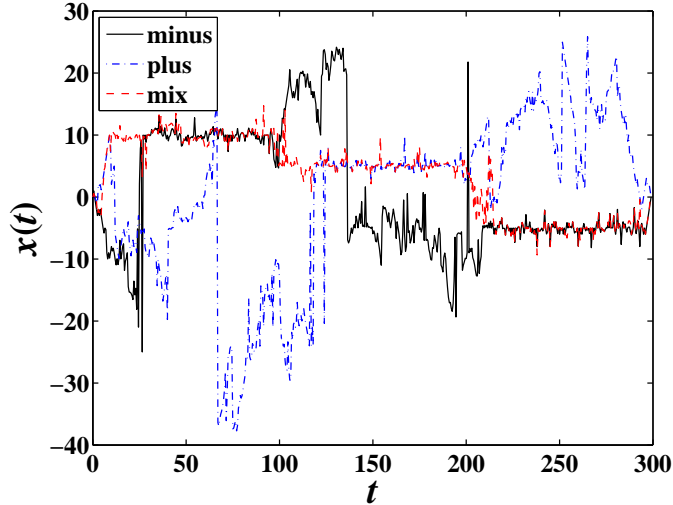}
\caption{(Color online) Comparison of the three lead-lag thermal
optimal paths using the three distances $\epsilon_-$ or
$\epsilon_+$, and $\epsilon_\pm$. The temperature is $T=0.1$.}
\label{Fig:TOP:Jump:cmp}
\end{figure}

Figure \ref{Fig:TOP:Jump:cmp} compares the reconstructed lead-lag
path $x(t)$ when using $\epsilon_-$ defined by (\ref{Eq:DM:minus}),
or $\epsilon_+$ defined by (\ref{Eq:DM:plus}), or $\epsilon_\pm$
defined by (\ref{Eq:DM:pm}). If the method worked perfectly, the
lead-lag path $x(t)$ would be equal to $x(t)=+10$ for $1\leqslant t
\leqslant 100$, $x(t)=+5$ for $101\leqslant t \leqslant 200$ and
$x(t)=-5$ for $201\leqslant t \leqslant 300$. One can observe that
the new proposed distance $\epsilon_\pm$ recovers the correct
solution up to moderate fluctuations. Unsurprisingly, the lead-lag
path reconstruction using $\epsilon_-$ gives the correct solution in
the first and third time intervals for which the correlation is
positive but is totally wrong with large fluctuations in the middle
time interval in which the correlation is negative. Symmetrically,
the lead-lag path reconstruction using $\epsilon_+$ gives the
correct solution in the middle interval where the correlation is
negative and is completely wrong with large fluctuations in the two
other intervals. Actually, we verify (not shown) that $\epsilon_\pm$
reduces to mostly $\epsilon_-$ in the first and third interval and
to $\epsilon_+$ in the middle interval, as it should.

\begin{figure}[htb]
\centering
\includegraphics[width=9cm]{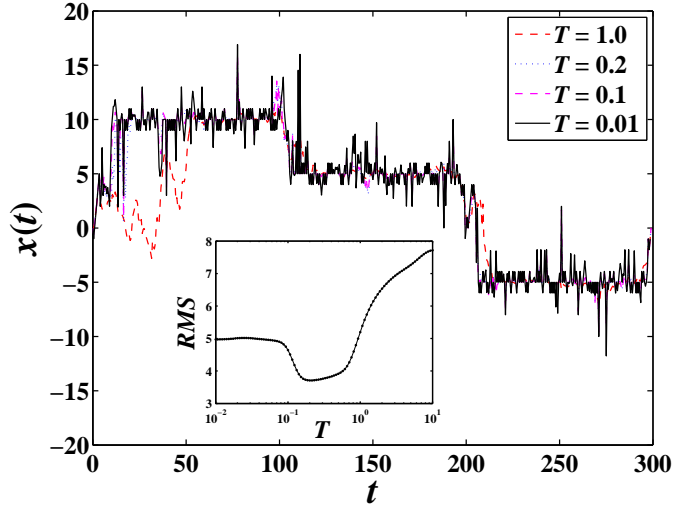}
\caption{(Color online) Thermally-averaged lead-lag paths of the
model (\ref{Eq:Jump}) for four different temperatures. Inset:
root-mean-square (rms) deviations between the reconstructed lead-lag
path and the exact solution ($x(t)=+10$ for $1\le t \le 100$,
$x(t)=+5$ for $101\le t \le 200$ and $x(t)=-5$ for $201\le t \le
300$) as a function of temperature in the range $0.01 \leq T \leq
10$.} \label{Fig:TOP:Jump:xt}
\end{figure}

Figure \ref{Fig:TOP:Jump:xt} tests the robustness of the
reconstructed lead-lag path using the distance $\epsilon_\pm$ with
respect to different choices of the temperature:  $T=1$, $0.2$,
$0.1$, and $0.01$. Recall that a vanishing temperature corresponds
to selecting the lead-lag path which has the minimum total sum of
distances along its length. At the opposite, a very large
temperature corresponds to wash out the information contained in the
distance matrix and treat all paths on the same footing. In between,
a finite temperature allows us to average the contribution over
neightboring paths with similar energies, making the estimated
lead-lag path more robust to noise-like structures in the distance
matrix due to noises decorating the two time series. It is apparent
that a too small temperature $T=0.01$ leads to spurious large spiky
fluctuations around the correct solution. A too large temperature
$T=1$ selects a thermally-averaged path which deviates from the
correct solution, here mostly at the beginning of the time series.
It seems that there is an optimal range of temperatures around
$T=0.1-0.2$ for which the correct solution is retrieved with minimal
fluctuations around it. The existence of an optimal range of
temperature is confirmed in the inset of Figure
\ref{Fig:TOP:Jump:xt}, which shows the root-mean-square (rms)
deviations between the reconstructed lead-lag path and the exact
solution ($x(t)=+10$ for $1\le t \le 100$, $x(t)=+5$ for $101\le t
\le 200$ and $x(t)=-5$ for $201\le t \le 300$) as a function of
temperature in the range $0.01 \leq T \leq 10$. The existence of a
well-defined optimal range of temperatures is strongest for smaller
signal-to-noise ratios $f\triangleq\sigma_\xi/\sigma_\eta$. For
large $f$ (weak noise), we observe that smaller temperatures are
better, as expected.

\begin{figure}[htb]
\centering
\includegraphics[width=9cm]{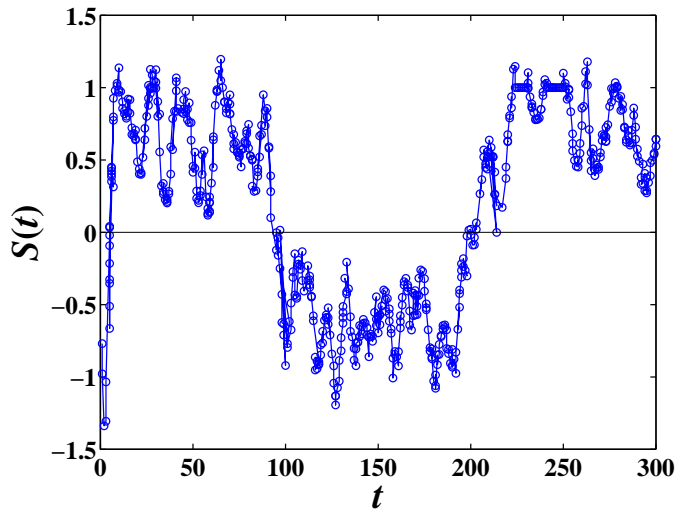}
\caption{Reconstruction of the sign of the cross-correlation of the
model (\ref{Eq:Jump},\ref{Eq:TOP:AR}) by the smoothed sign
recognition given by expression (\ref{Eq:Sign}).}
\label{Fig:TOP:Jump:Signal}
\end{figure}

The whole purpose of the new distance $\epsilon_\pm$ is to be able
to identify, not only the lead-lag structure better but also, the
existence of possible negative correlations as well as changes of
the sign of the correlation with time. We identify the sign
$s(t,x(t)) = s(t_1,t_2)$ of the cross-correlation of the two time
series at the times $t_1,t_2$ from the value of $\epsilon_\pm$: when
$\epsilon_\pm$ reduces to $\epsilon_-$ (resp. $\epsilon_+$), we
conclude that the correlation is positive (resp. negative). The
corresponding algorithm for the sign of the cross-correlations is
thus
\begin{equation}\label{Eq:Sign}
    s(t) = s(t_1,t_2) = \left\{
    \begin{array}{cc}
      +1 & ~~{\rm{if}}~~ \epsilon_\pm=\epsilon_- \\
      -1 & ~~{\rm{if}}~~ \epsilon_\pm=\epsilon_+
    \end{array}
    \right.
\end{equation}
Due to the noises on the two time series, $s(t)$ is also noisy. Thus,
to obtain a meaningful information on the sign of the
cross-correlations, we apply a smoothing algorithm to $s(t)$. For
this, we use the Savitzky-Golay filter with a linear function and
include 21 points to the left of each time (to ensure causality).
The filtered signal $S(t)$ is shown in
Fig.~\ref{Fig:TOP:Jump:Signal}. The results are quite consistent
with the model in which the correlation is negative in the middle
period $100<t<200$ and positive otherwise.

\section{Historical volatilities of inflation rate and economic output rate}
\label{s1:Appl}

In this section, we apply our novel technique to the relationship
between inflation and real economic output quantified by GDP in the
hope of providing new insights. This problem has attracted
tremendous interests in past decades in the macroeconomic
literature. Different theories have suggested that the impact of
inflation on the real economy activity could be either neutral,
negative, or positive. Based on the story of Mundell that higher
inflation would lower real output \cite{Mundell-1963-JPE}, Tobin
argued that higher inflation causes a shift from money to capital
investment and raise output per capita \cite{Tobin-1965-Em}, known
as the Mundell-Tobin effect. On the contrary, Fischer suggested a
negative effect, stating that higher inflation resulted in a shift
from money to other assets and reduced the efficiency of
transactions in the economy due to higher search costs and lower
productivity \cite{Fischer-1974-EI}. In the middle ground, Sidrauski
proposed a neutral effect where exogenous time preference fixed the
long-run real interest rate and capital intensity
\cite{Sidrauski-1967-AER}. These arguments are based on the rather
restrictive assumption that the Philips curve (inverse relationship
between inflation and unemployment), taken in addition to be linear,
is valid. To evaluate which model characterizes better real economic
systems, numerous empirical efforts have been performed and the
question is still open.

On the other hand, much focus is put on the nexus between inflation
and its uncertainty and economic activity. Okun made the hypothesis
of a positive correlation between inflation and inflation
uncertainty \cite{Okun-1971-BPEA}. Furthermore, Friedman argued that
an increase in the uncertainty of future inflation reduces the
economic efficiency and lowers the real output rate
\cite{Friedman-1977-JPE}, which is verified empirically (see, e.g.
\cite{Davis-Kanago-1996-OEP,Davis-Kanago-1998-JMCB,AlMarhubi-1998-AE,Grier-Perry-2000-JAEm,Hayford-2000-JMe,Fountas-Karanasos-Kim-2006-OBES}).
Following the seminal work of Taylor \cite{Taylor-1979-Em}, the
output-inflation variability trade-off has been tested extensively
in the literature, such as in
\cite{Defina-Stark-Taylor-1996-JMe,Fuhrer-1997-JMCB,Cobham-Macmillan-Mcmillan-2004-AEL,Lee-2002-SEJ,Lee-2004-CEP},
which are based on model specification. Liu and Liu analyzed the
relation between the historical volatility of the output and of the
inflation rate, using Chinese data from 1992 to 2004
\cite{Liu-Liu-2005-ERJ}. They found that there is a strong
correlation between the two volatilities and, what is more
interesting, that the rolling correlation coefficient changes its
sign. In the following, we investigate the nexus between the
historical volatilities of inflation and output in a model-free
manner to test for possible changes of the signs of their
cross-correlation structure.

\begin{figure}[htb]
\centering
\includegraphics[width=9cm]{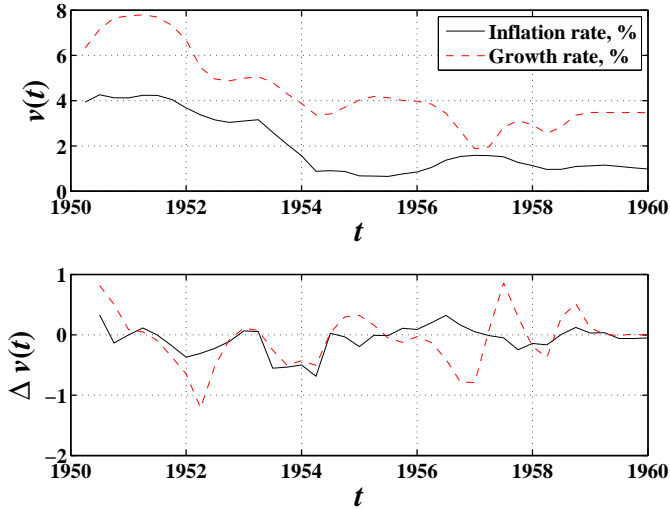}
\caption{Upper panel: quarterly historical volatilities of the
annualized inflation rate and economic growth rate of the United
States of America; lower panel: their quarterly changes.}
\label{Fig:TOP:InfGDP:VIVG}
\end{figure}

The data sets, which were retrieved from the FRED II database,
include monthly consumer price index (CPI) for all urban consumers
and seasonally adjusted quarterly gross domestic product (GDP)
covering the time period from 1947 to 2005. The annualized rates of
inflation rate $r_{\rm{CPI}}$ and economic growth rate
$r_{\rm{GDP}}$ were calculated on a quarterly basis from the CPI and
GDP respectively. The historical volatility is calculated in a
rolling window as
\begin{equation}\label{Eq:TOP:VIVG}
    \nu(t) = \left[\frac{1}{\Delta{t}}\sum_{s=t-\Delta{t}+1/4}^{t} \left[r(t)-\mu(t)\right]^2
    \right]^{1/2}~,
\end{equation}
where $r=r_{\rm{CPI}}$ for inflation rate and $r=r_{\rm{GDP}}$ for
growth rate, and $\mu(t)$ is their corresponding mean in the rolling
window $[t-\Delta{t}+1/4,t]$. The unit of $t$ and $\Delta{t}$ is one
year. The resulting historical volatility series $\nu_{\rm{CPI}}(t)$
and $\nu_{\rm{GPD}}(t)$ are shown in the upper panel of
Fig.~\ref{Fig:TOP:InfGDP:VIVG} for the time period $[1950,1960]$,
with $\Delta{t}=3$ years. Since the volatility $\nu(t)$ is
non-stationary (as shown by a standard unit-root test), we use the
first-difference of volatility $\Delta{\nu}(t)$, shown in the lower
panel of Fig.~\ref{Fig:TOP:InfGDP:VIVG}. We focus on the 10-year
time period $[1950,1960]$ only for a clearer visualization since the
analysis and results are the same qualitatively in other time
periods.

\begin{figure}[htb]
\centering
\includegraphics[width=9cm]{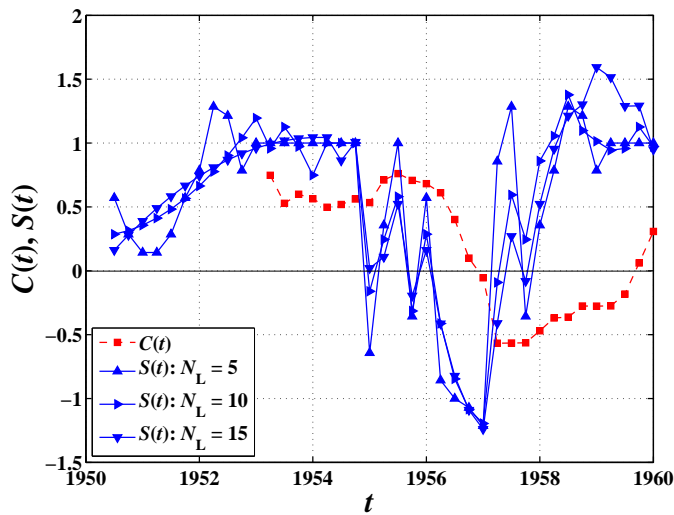}
\caption{Determination of the sign of the correlation between the
variations of the volatilities $\nu_{\rm{CPI}}(t)$ and
$\nu_{\rm{GPD}}(t)$ as a function of time in a running window of
three years. Our new method $S(t)$ (triangles with three values of
the smoothing parameter $N_L$) is compared with the
cross-correlation $C(t)$ in a running window of size equal to three
years (squares).} \label{Fig:TOP:InfGDP:Convention}
\end{figure}

Visual inspection of the lower panel of
Fig.~\ref{Fig:TOP:InfGDP:VIVG} suggests that the variations of the
volatilities $\nu_{\rm{CPI}}(t)$ and $\nu_{\rm{GPD}}(t)$ are
approximately synchronous from 1951 to 1954 and then become
approximately anti-phased from 1955 to 1958. Can this be confirmed
or falsified by the technique proposed here? To address this
question, we determine the smoothed sign function $S(t)$ determined
as explained at the end of the previous section. Our tests show that
the lead-lag path is close to the diagonal and that there is no
significant gain obtained by allowing for a time-varying lag between
the variations of the volatilities $\nu_{\rm{CPI}}(t)$ and
$\nu_{\rm{GPD}}(t)$. We thus calculate $S(t)$ by smoothing the
signal $s(t)$ defined by (\ref{Eq:Sign}) with the distance matrix
constructed using definition (\ref{Eq:DM:pm}) along the diagonal of
the plane $(t_1,t_2)$ (in other words, for $x(t)=0$). We again use
the causal Savitzky-Golay filter with a quadratic polynomial and
$N_L$ data points to the left of each time step $t$ plus the point
at $t$ itself. As shown in Fig.~\ref{Fig:TOP:InfGDP:Convention}, we
find that the sign signal function $S(t)$ is quite robust with
respect to variations of the smoothing parameter $N_L$ in the range
$N_L=5-15$. For comparison, we also plot in
Fig.~\ref{Fig:TOP:InfGDP:Convention} the cross-correlation function
$C(t)$ in rolling windows of three years.

The reconstructed sign of the correlations between variations of the
volatilities $\nu_{\rm{CPI}}(t)$ and $\nu_{\rm{GPD}}(t)$ is in good
agreement with and actually makes more precise the visual impression
mentioned above. In particular, one can observe that the transition
from a synchronicity to anti-phased was gradual with possible ups
and downs before the anti-correlation set in in 1956. In contrast,
the cross-correlation method suffers from a serious lack of
reactivity, predicting a change of correlation sign two years or so
after it actually happened. We can thus conclude that our new
measure outperforms significantly the traditional cross-correlation
measure for real-time identification of switching of correlation
structures.

\section{Concluding remarks}
\label{s1:concl}

We have extended the thermal optimal path method
\cite{Sornette-Zhou-2005-QF,Zhou-Sornette-2006-JMe} in order to, not
only identify the time-varying lead-lag structure between two time
series but also, to measure the sign of their cross-correlation. In
so doing, the identification of the lead-lag structure is improved
when there is the possibility for the sign of their correlation to
shift. In this goal, the main modification of the method previously
introduced in
Refs.\cite{Sornette-Zhou-2005-QF,Zhou-Sornette-2006-JMe} consists in
generalizing the distance matrix in such a way that both correlated
and anti-correlated time series can be matched optimally.

A synthetic numerical example has been presented to verify the
validity of the new method. Extensive numerical simulations have
determined the existence of an optimal range $T\sim(0.1,1)$ of
temperatures to use for the robust thermal averaging. We have also
proposed a new measure, the sign signal function $S(t)$, that allows
us to identify the sign of the correlation structure between two
time series.

We have applied our new method to the investigation of possible
shifts between synchronous to anti-phased variations of the
historical volatility of the USA inflation rate and economic growth
rate. The two variables are found positively correlated and in a
synchronous state in the 1950's except over the time period from the
last quarter of 1954 till around 1958, when they were in a
asynchronous phase (approximately anti-phased). While the
traditional cross-correlation function fails to capture this
behavior, our new TOP method provides a precise quantification of
these regime shifts.

The emphasis of this paper has been methodological. Extensions will
investigate the economic meaning of the change of correlation
structures as shown here. One possible candidate is the concept of
shifts of convention, as discussed in the introduction. More work on
many more examples is needed to ascertain the generality of these
effects. Overall, the development of better and more precise
quantitative tools is progressively unraveling a picture according
to which variability and changes of correlation structures is the
rule rather than the exceptions in macroeconomics and in financial economics,
in the spirit of Aoki and Yoshikawa \cite{Aoki-Yoshikawa-2006}.

\bigskip
{\textbf{Acknowledgments:}}

We are grateful to M. Wyart for helpful discussions. This work was
partially supported by the National Natural Science Foundation of
China (Grant No. 70501011), the Fok Ying Tong Education Foundation
(Grant No. 101086), and the Alfred Kastler Foundation.

\bibliography{E:/papers/Bibliography}

\end{document}